\documentstyle[aaspp4,epsf,astrobib,11pt]{article}

\input{epsf}

\let\oldfootsep=\footnotesep
\def\msun { \rm {{\em M}_\odot}} 
\def\rsun { \rm {{\em R}_\odot}}

\def\umin{u_{\rm min}}
\def\rstar{R_{\rm *}} 
 
\def\tstar{t_{\rm *}}

\def\vperp{v_{\rm \perp}}
\def\umin{u_{\rm min}}
\def\t0{t_{\rm 0}}

\def\vs {{\bf v}_S} 
\def\vl {{\bf v}_L} 
\def\vhbold {\widehat{\bf v}} 
\def\thsky {\alpha}  
\def\vhat{\widehat{v}} 
\def\that {\widehat{t}}
 
\def\kms {\,{\rm km \, s^{-1} }}
\def\kpc {\, {\rm kpc}}
\def\spose#1{\hbox to 0pt{#1\hss}}
\def\simlt{\mathrel{\spose{\lower 3pt\hbox{$\mathchar"218$}}
     \raise 2.0pt\hbox{$\mathchar"13C$}}}
\def\simgt{\mathrel{\spose{\lower 3pt\hbox{$\mathchar"218$}}
     \raise 2.0pt\hbox{$\mathchar"13E$}}}

\begin{document}

\title{Discovery and Characterization of a Caustic Crossing 
       Microlensing Event in the SMC}
\author{
  C.~Alcock\altaffilmark{1,2},           
  R.A.~Allsman\altaffilmark{3},          
  D.~Alves\altaffilmark{1,4},            
  T.S.~Axelrod\altaffilmark{5},          
  A.C.~Becker\altaffilmark{2,6},         
  D.P.~Bennett\altaffilmark{1,2,7},      
  K.H.~Cook\altaffilmark{1},             
  A.J.~Drake\altaffilmark{5},            
  K.C.~Freeman\altaffilmark{6},          
  K.~Griest\altaffilmark{2,8},           
  L.J.~King\altaffilmark{2,7},           
  M.J.~Lehner\altaffilmark{9},           
  S.L.~Marshall\altaffilmark{1},         
  D.~Minniti\altaffilmark{1,10},         
  B.A.~Peterson\altaffilmark{6},         
  M.R.~Pratt\altaffilmark{11},           
  P.J.~Quinn\altaffilmark{12},           
  S.H.~Rhie\altaffilmark{7},             
  A.W.~Rodgers\altaffilmark{5},          
  P.B.~Stetson\altaffilmark{13},         
  C.W.~Stubbs\altaffilmark{2,6},         
  W.~Sutherland\altaffilmark{14},        
  A.~Tomaney\altaffilmark{6},            
  T.~Vandehei\altaffilmark{2,8},         
\begin{center}
{\bf (The MACHO/GMAN Collaboration) }\\
\end{center}
}

\altaffiltext{1}{Lawrence Livermore National Laboratory, Livermore, CA 94550
}

\altaffiltext{2}{Center for Particle Astrophysics,
  University of California, Berkeley, CA 94720
}

\altaffiltext{3}{Supercomputing Facility, Australian National University,
  Canberra, ACT 0200, Australia 
}
 
\altaffiltext{4}{Department of Physics, University of California, Davis, CA 95616
}

\altaffiltext{5}{Mt.~Stromlo and Siding Spring Observatories,
  Australian National University, Weston, ACT 2611, Australia
}
 
\altaffiltext{6}{Departments of Astronomy and Physics,
  University of Washington, Seattle, WA 98195
}
 
\altaffiltext{7}{Department of Physics, University of Notre Dame, Notre Dame, IN 46556
}

\altaffiltext{8}{Department of Physics, University of California,
  San Diego, CA 92093
}
 
\altaffiltext{9}{Department of Physics, University of Sheffield,
Sheffield s3 7RH, UK
}

\altaffiltext{10}{Departmento de Astronomia, P. Universidad Cat\'olica, 
Casilla 104, Santiago 22, Chile
} 

\altaffiltext{11}{Center for Space Research, MIT,
Cambridge, MA 02139
}
 
\altaffiltext{12}{European Southern Observatory, Karl-Schwarzchild Str. 2, D-857
48, Garching, Germany
}
 
\altaffiltext{13}{National Research Council,
  5071 West Saanich Road, RR 5, Victoria, BC V8X 4M6, Canada
}
 
\altaffiltext{14}{Department of Physics, University of Oxford,
  Oxford OX1 3RH, U.K.
}
 
 
\setlength{\footnotesep}{\oldfootsep}
\renewcommand{\topfraction}{1.0}
\renewcommand{\bottomfraction}{1.0}     


\vspace{-5mm}
\begin{abstract} 
\rightskip = 0.0in plus 1em

We present photometric observations and analysis of the second
microlensing event detected towards the Small Magellanic Cloud (SMC),
MACHO Alert 98-SMC-1.  This event was detected early enough to allow
intensive observation of the lightcurve.  These observations revealed
98-SMC-1 to be the first caustic crossing, binary microlensing event
towards the Magellanic Clouds to be discovered in progress.
 
Frequent coverage of the evolving lightcurve allowed an accurate
prediction for the date of the source crossing out of the lens caustic
structure.  The caustic crossing temporal width, along with the
angular size of the source star, measures the proper motion of the
lens with respect to the source, and thus allows an estimate of the
location of the lens.  Lenses located in the Galactic halo would have
a velocity projected to the SMC of $\vhat \sim 1500 \kms$, while an
SMC lens would typically have $\vhat \sim 60 \kms$.  The event
lightcurve allows us to obtain a unique fit to the parameters of the
binary lens, and to estimate the proper motion of the lensing system.

We have performed a joint fit to the MACHO/GMAN data presented here,
including recent EROS data of this event \cite{eros-98smc1}.  These
joint data are sufficient to constrain the time $\tstar$ for the lens
to move an angle equal to the source angular radius; $\tstar = 0.116
\pm 0.010$ days.  We estimate a radius for the lensed source of
$\rstar = 1.4 \pm 0.1 \rsun$ from its unblended color and magnitude.
This yields a projected velocity of $\vhat = 84 \pm 9 \kms$.  Only
0.15 \% of halo lenses would be expected to have a $\vhat$ value at
least as small as this, while 31\% of SMC lenses would be expected to
have $\vhat$ as large as this.  This implies that the lensing system
is more likely to reside in the SMC than in the Galactic halo.
Similar observations of future Magellanic Cloud microlensing events
will help to determine the contribution of Machos to the Galaxy's dark
halo.

\end{abstract}
\vspace{-5mm}

\newpage
\section{Introduction}
\label{sec-intro}

The MACHO project is monitoring stars in the Large Magellanic Cloud
(LMC), in the Small Magellanic Cloud (SMC) and in the Galactic center,
searching for the transient brightening that is characteristic of
gravitational microlensing (e.g. \citeNP{paczynski86}). Our survey has
detected over 250 likely lensing events to date. The majority of these
are seen towards the Galactic center, presumably arising from lensing
due to stars in the disk and bulge of the Galaxy.  The event rates
towards the LMC and SMC can be used to ascertain whether astrophysical
objects comprise the dark matter halo of the Milky Way
(\citeNP{roulet97,pac-rev}).  The rate of microlensing seen towards
the LMC exceeds that predicted from known Galactic sources
\cite{macho-apj97d}. The nature of this lensing population is unknown.

Are the excess LMC microlensing events due to Galactic dark matter in
the form of Machos?  The key to resolving this issue lies in
determining the location of the lensing objects.  If the excess
microlensing optical depth seen towards the Magellanic Clouds is due
to lensing by objects in the Magellanic Clouds, then the microlensing
surveys will have ruled out Machos in the mass range
$10^{-7}$-$1\msun$ as significant contributors to the mass of the
Galaxy's dark halo \cite{macho-eros,macho-apj97d,eros-lmc}.
 
An important development in the microlensing field has been the
community's ability to detect microlensing events in real time,
allowing concentrated photometric and spectroscopic follow--up
observations
(\citeNP{macho-apj97e,planet95,udalski94d,stubbs-realtime}). The
survey projects alert the community to ongoing microlensing events,
and maintain web sites with pertinent information~\footnote[1]{
 
EROS~~~~~{\bf~http://www-dapnia.cea.fr/Spp/Experiences/EROS/alertes.html}
 
MACHO~~{\bf~http://darkstar.astro.washington.edu}
 
OGLE~~~~~{\bf~http://www.astrouw.edu.pl/~ftp/ogle/ogle2/ews/ews.html} }.

Occasionally a microlensing event's light curve is seen to deviate
from the characteristic shape expected for a point--like lens in
inertial motion relative to the line of sight from us to a point--like
source star.  These ``exotic'' lensing events can break the degeneracy
between the lens' mass, location, and transverse velocity inherent in
the standard gravitational microlensing model.  In particular, if the
foreground lens is a binary system, it is likely that the source will
cross a lens ``caustic'' where two new images are created with a
magnification which is singular for an instant, in the point source
limit (\citeNP{dominik98a,rhie,mao-pac}).  Once a caustic crossing has
occurred, it is certain that a second caustic crossing will also occur
because the extra images must disappear before the source departs the
lens region.
 
Caustic crossing events, if adequately monitored, provide an
opportunity to measure how long it takes for the caustic line to
transit the face of the source star.  Given the angular size of the
source star, this provides a measurement of the transverse angular
speed of the binary lens with respect to the source, which constrains
the lens' transverse linear speed as a function of its distance along
the line of sight. Such measurements are also possible for single lens
events such as MACHO 95-BLG-30 \cite{macho-apj97e}, but finite source
effects from single lenses will be extremely rare for sources of small
angular size like those in the Magellanic Clouds.
 
\section{98-SMC-1 : Discovery and Observations }
\label{sec-alert}

Our ongoing survey program uses the 1.3m Great Melbourne telescope at
Mt. Stromlo to produce dual--color photometry
(\citeNP{macho-hart96a,macho-marshall94,macho-stubbs93a}). In
addition, we have been granted the use of roughly one hour per night
on the CTIO~0.9m telescope as service observing. We also make use of
the 0.8m Reynolds telescope at Mt. Stromlo for following ongoing
lensing events.
 
The source star in MACHO Alert 98-SMC-1 is located at $\alpha =
$00:45:35.2, $\delta = -$72:52:34.1 (J2000).  A finding chart is
available from the authors.  MACHO Alert 98-SMC-1 was announced May
25.9 UT
after the source had apparently brightened by $\sim 0.9$ mag. The
baseline magnitude and color of the MACHO object had been constant
over 5 years of observations, at $V = 21.37 \pm 0.10, V-R = 0.14 \pm
0.05$.  Our binary microlens fit presented in
Section~\ref{sec-analysis} indicates that the lensed source has a
brightness of $V = 22.05 \pm 0.15, V-R = 0.03 \pm 0.10$, with the
additional unlensed flux provided by other sources within the seeing
disk (as quite commonly occurs in these crowded fields).

Follow--up observations were scheduled nightly on the CTIO 0.9m
telescope, which showed a gradual rise in the lightcurve until
June 6.5 UT, when it was noticed the source had brightened suddenly by
1.5 mag.  Photometry from the subsequent night confirmed that this was
a likely caustic crossing event, and an IAU Circular was submitted
alerting the astronomical community to the first real--time detection
of a binary lensing event towards the Magellanic Clouds
\cite{macho-iauc98a}.
 
Continued observations allowed an estimate on the date of the second
caustic crossing of June $19.3 \pm 1.5$ UT \cite{macho-iauc98b} to be
issued June 15.3. This error bar was estimated based upon prior
experience predicting caustic crossings (see \citeNP{macho-iauc96}),
and also because the path of the source was expected to cross the
caustic at a small angle, so that a small error in the source
trajectory translated into a large error in the time of the caustic
crossing. Data from June 15 caused a revision of the caustic crossing
time to June $19.2 \pm 1.5$ UT, and June 17 data forced a revised
prediction of June 18.2.

\begin{figure}
\plotone{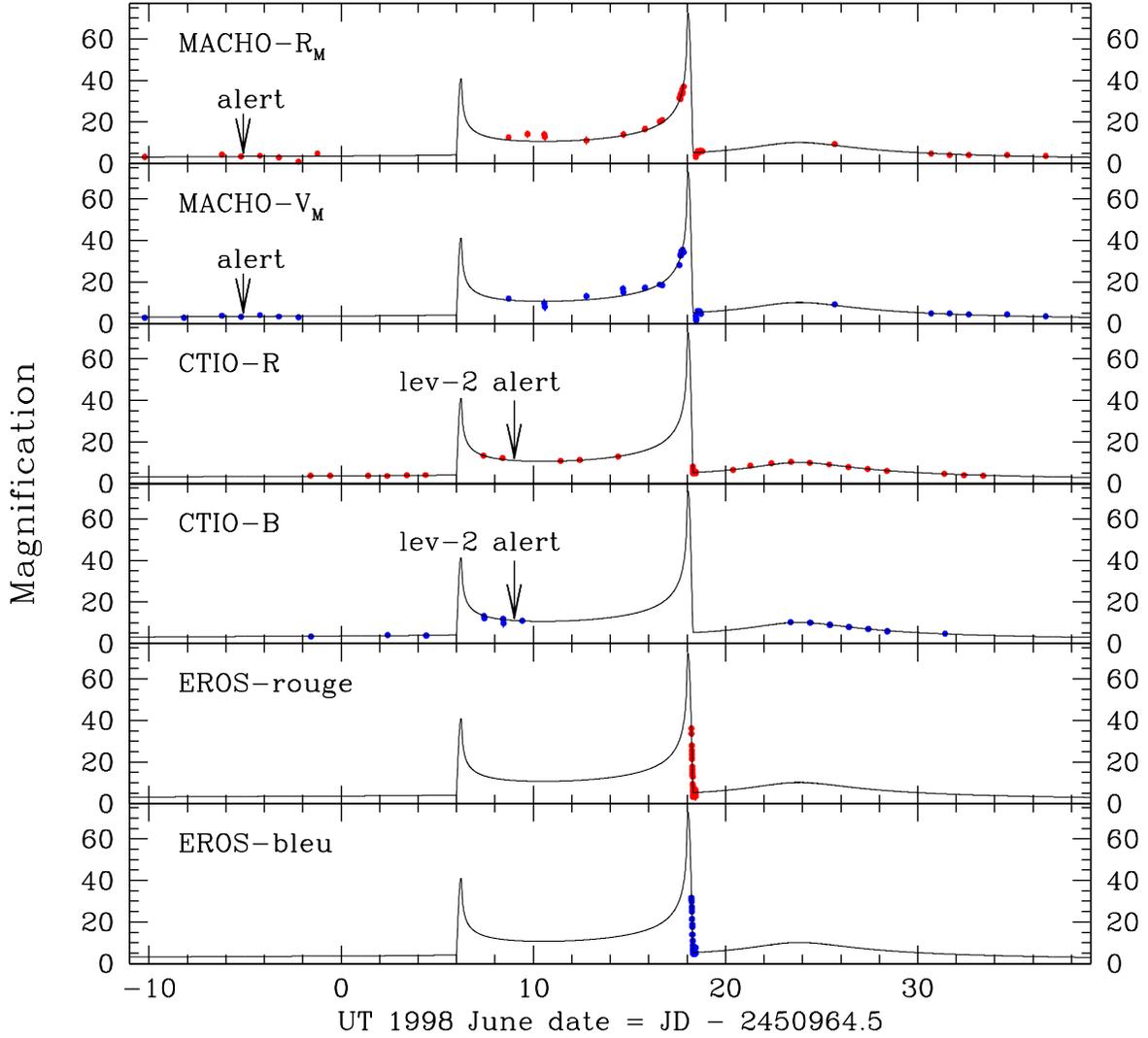}
\figcaption[f1.ps]{\label{fig-lc1}
The light curve of event 98-SMC-1. The panels show magnification as a
function of time, with passbands and sites as indicated.  The times of
the initial alert and the confirmation of the caustic crossing are
shown with arrows. The best fit binary microlensing curve is shown as
a solid line. }
\end{figure}

Figure~\ref{fig-lc1} shows the joint MACHO/GMAN/EROS light curve for
98-SMC-1~\footnote[2]{ MACHO/GMAN data for this event are available at
ftp://darkstar.astro.washington.edu/macho/Alert/98-SMC-1/ } : dual
color lightcurves from the MACHO Project's Mt.~Stromlo 1.3m telescope,
the CTIO 0.9m telescope, and the EROS Project's 1.0m Marly telescope
at La Silla \cite{palanque-thesis}.  The CTIO lightcurves are for the
standard R and B bands, while the MACHO and EROS lightcurves are for
their respective non-standard passbands.  The MACHO and EROS data were
reduced with the SoDOPHOT photometry package used for routine
reduction of MACHO data, while the CTIO data were reduced with
ALLFRAME \cite{stetson-allframe}.  In each case, the photometric error
estimates reported by the photometry code are used, with minimum
errors of 0.014, 0.01, and 0.005 added in quadrature for the MACHO,
EROS, and CTIO data, respectively.  The CTIO error estimates are also
multiplied by 1.5, although this appears to overestimate the errors
for this event.  Each lightcurve is normalized to the best fit
unlensed flux of the source, so that the brightness of any unlensed
(blended) sources has been subtracted out.  The overall structure is
consistent with a caustic crossing binary lens event.

\begin{figure}
\plotone{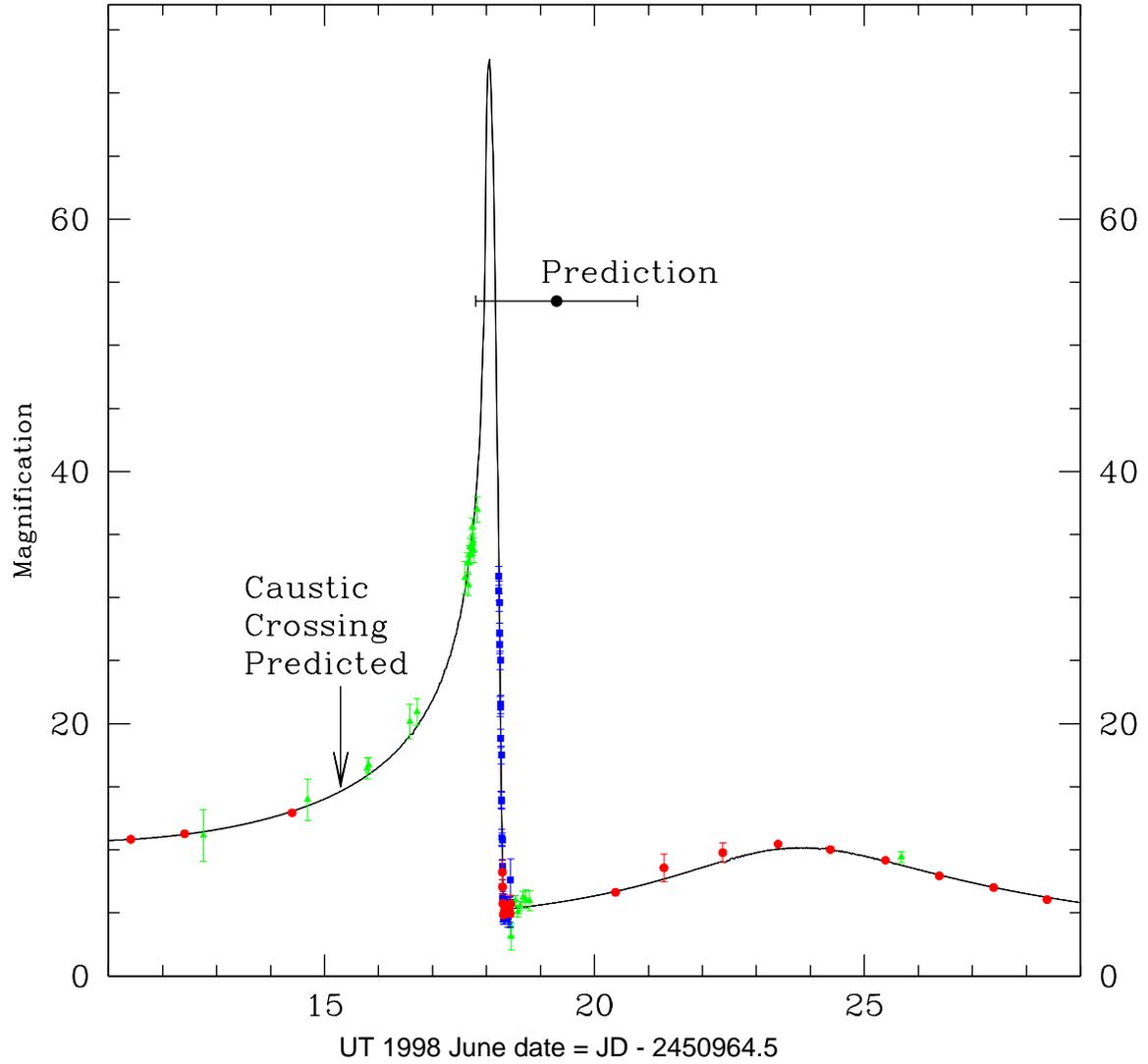}
\figcaption[f2.ps]{\label{fig-lc2}
A close-up view of the last half of the light curve of event 98-SMC-1.
Included are the date of the original second caustic crossing
prediction, and the prediction with error bars of 1.5 days.  Only the
MACHO-R (green triangles), CTIO-R (red circles) and EROS-B (blue squares)
data are plotted.}
\end{figure}

Figure~\ref{fig-lc2} shows a close-up of the latter part of the
lightcurve for the MACHO-R, CTIO-R, and EROS-B data.  The CTIO-R band
data were generally taken over a period of less than 1.5 hours, and
have been binned nightly for every night except for June 18.

\section{Analysis}
\label{sec-analysis}
 
Once it was realized that 98-SMC-1 was a likely binary microlensing
event, it was modeled using the binary lens fitting code developed by
two of us \cite{benn-rhie98,benn-rhie96}.  Starting with the data
available on June 8, we began a series of Monte Carlo searches that
initially resulted in several possible binary microlensing fits. By
June 14, only one good fit remained, and an accurate caustic crossing
prediction was announced \cite{macho-iauc98b}.  This preliminary
lightcurve and fit parameters were posted on the WWW~\footnote[3]{
 
see~~{\bf http://darkstar.astro.washington.edu/98-SMC-1\_lev2.html} 

and~~{\bf http://bustard.phys.nd.edu/MPS/98-SMC-1/} }.

A caustic crossing binary lens lightcurve can be described by 7
parameters if the orbital motion of the lensing objects is
neglected. These parameters include the 3 parameters for a single lens
fit which are $\that$, the Einstein diameter crossing time; $\umin$,
the distance of closest approach between the lens center of mass and
the source; and $t_0$, the time of the closest approach.  The 3
intrinsic binary parameters are the binary source separation, $a$, the
mass fraction of the mass \# 1, $\epsilon_1$, and the angle between
the lens axis and the source trajectory, $\theta$.  The final
parameter is the stellar radius crossing time $\tstar$, which is the
time for the lens to move relative to the source by an angle equal to
the source angular radius.  Thus the caustic crossing duration is $2
\tstar / \sin \phi$, where $\phi$ is the angle between the relative
motion vector and the caustic line.  This parameter is sensitive to
the assumed limb darkening model.  We use a simple linear limb
darkening model \cite{claret-limbd} with coefficients 0.482, 0.620,
0.506, 0.697, 0.460, and 0.561 for the MACHO-R, MACHO-V, CTIO-R,
CTIO-B, EROS-R, and EROS-B passbands, respectively.

Table~\ref{tab-fpar} shows the June 15 pre-caustic crossing fit
parameter estimates (as the first line) and the current best fit
parameters with errors in the second line.  Note that the parameters
have changed only slightly between the pre-caustic crossing fit and
the current fit.  The fit lightcurves are shown in
Figures~\ref{fig-lc1} and~\ref{fig-lc2}, and the parameters of the fit
are shown in Tables~\ref{tab-fpar} and~\ref{tab-bpar}.  Fit
statistics are shown in Table~\ref{tab-fstat}.  Note that the $\chi^2$
per degree of freedom is less than 1 for every passband, except for
the MACHO passbands where it is somewhat larger than 1.  The excess
fit $\chi^2$ in the MACHO data is partially due to excess scatter in
the baseline, and from very high airmass observations on June 18.
However, there is a significant ``bump'' in the MACHO lightcurves near
June 17.7 which is not easily explained away as photometric
error. This might be consistent with a caustic crossing of a binary
companion 3-4 magnitudes fainter than the primary lensed star.

However, assuming a single source, the uniqueness of this fit seems
sound.  The MACHO data characterize the full lightcurve.  Each night
of CTIO observations, begun soon after MACHO's initial alert,
constrains the overall lightcurve at the $\lesssim 3\%$ level.
Especially important constraints are the flux preceding the first
caustic crossing, the minimum flux between caustics, and the magnitude
of the cusp passage after the second caustic crossing, which strongly
constrains $\theta$ and avoids fit uncertainties present in other
analyses of this event (\citeNP{planet-98smc1,eros-98smc1}).  The two
fits presented by the PLANET collaboration \cite{planet-98smc1}, for
example, both appear inconsistent with the pre-caustic crossing MACHO
and CTIO data.  However, PLANET was able to densely sample the peak of
the caustic crossing, measuring to high precision the time taken by
the caustic to transit the source.  The EROS data used in this
analysis strongly constrain the falling slope and end-point of the
caustic crossing, which leads to our estimate of the lens velocity
projected to the SMC.

\subsection{Projected Lens Velocity ($\vhat$)}
\label{sec-vhat}

In order to convert the fit parameter $\tstar$ to a projected velocity
$\vhat$, we need to estimate the radius of the source star. The color
of the lensed star is $V-R = 0.03$, which may be determined from the
microlensing fit or simply from the mean color of the MACHO
observations on June 17, when the entire star was highly magnified.
Assuming $E(V-R) = 0.03$ and $T_{\rm eff} = 8000 K$
\cite{planet-98smc1}, we obtain $\rstar = 1.4 \pm 0.1 \rsun$ and $M_V
= 2.8 \pm 0.3$ \cite{bertelli,lang}.  For an assumed source distance
of $60\kpc$, this yields $m_V = 21.8 \pm 0.3$, which is consistent
with the value $m_V = 22.05 \pm 0.15$ determined from the microlensing
fit and the MACHO photometric zero point determination.

If we divide the stellar radius of $\rstar = 1.4 \pm 0.1 \rsun$
by the stellar radius crossing time $\tstar = 0.116 \pm 0.010$ days, 
we obtain a projected velocity of $\vhat = 84 \pm 9 \kms$, which can be
used to estimate the location of the lens.

By definition, $\vhat \equiv v_\perp / x$, where $v_\perp$ is the
tangential velocity of the lens relative to the observer-source line,
and $x$ is the ratio of the observer-source to observer-lens distances
(cf \citeN{han-gould96}).  For a given lens mass, the rate of
microlensing is proportional to
\begin{equation}
\label{eq-gamma}
d\Gamma \propto \sqrt{x (1-x)} \, \rho_L (x) \, v_\perp f_S(\vs) \,
f_L(\vl) \, dx \, d\vs \, d\vl .
\end{equation}
where $\rho_L$ is the density of lenses at distance $x$, $f_L(\vl)$
and $f_S(\vs)$ are the 2-D lens and source velocity distribution
functions (normalized to unity) in the plane perpendicular to the line
of sight.  The source and lens velocities $\vs, \vl$ are related to
$\vhbold$ by $\vl = (1-x) {\bf v}_\odot + x (\vs + \vhbold)$, where
$\vhbold = (\vhat \cos \thsky, \vhat \sin \thsky)$ and $\thsky$ is the
(unknown) direction on the sky of the relative proper motion.

Given a model for $\rho_L, f_S, f_L$, we may integrate
eq.~\ref{eq-gamma} and thus obtain joint probability distributions for
any of the variables.  We need to consider the joint probability
distribution of events in the $(x, \vhat)$ plane, and then integrate
over $x$ to get a probability distribution of $\vhat$ for each lens
population.  This gives
\begin{equation} 
\label{eq-like}
 {\cal L}(\vhat) = \left.{d\Gamma \over d\vhat}\right|_{\vhat}  \propto 
 \int \, dx \; 
 \sqrt{x (1-x)} \, \rho_L (x) \, \vhat^2  \, x^3 
 \int \,  d\vs \, d\thsky \; f_S(\vs) 
 \, f_L\left( (1-x) {\bf v}_\odot + x (\vs + \vhbold ) \right) 
\end{equation}

We evaluate eq.~\ref{eq-like} over $\vhat$ separately for lenses
either in the dark halo or the SMC. We assume that the source is in
the SMC, thus $f_S$ is a Gaussian with velocity dispersion of $30
\kms$ in each component, centered on the SMC mean velocity which is
taken as the projection of $(U,V,W) = (40,-185,171) \kms$
\cite{gardiner} onto the plane normal to the line of sight.  For halo
lenses we assume $\rho_L$ is a `standard' isothermal sphere with core
radius 5 kpc. We take $f_L$ to be a Gaussian with zero mean and
velocity dispersion $155 \kms$ in each component.

\begin{figure}
\plotone{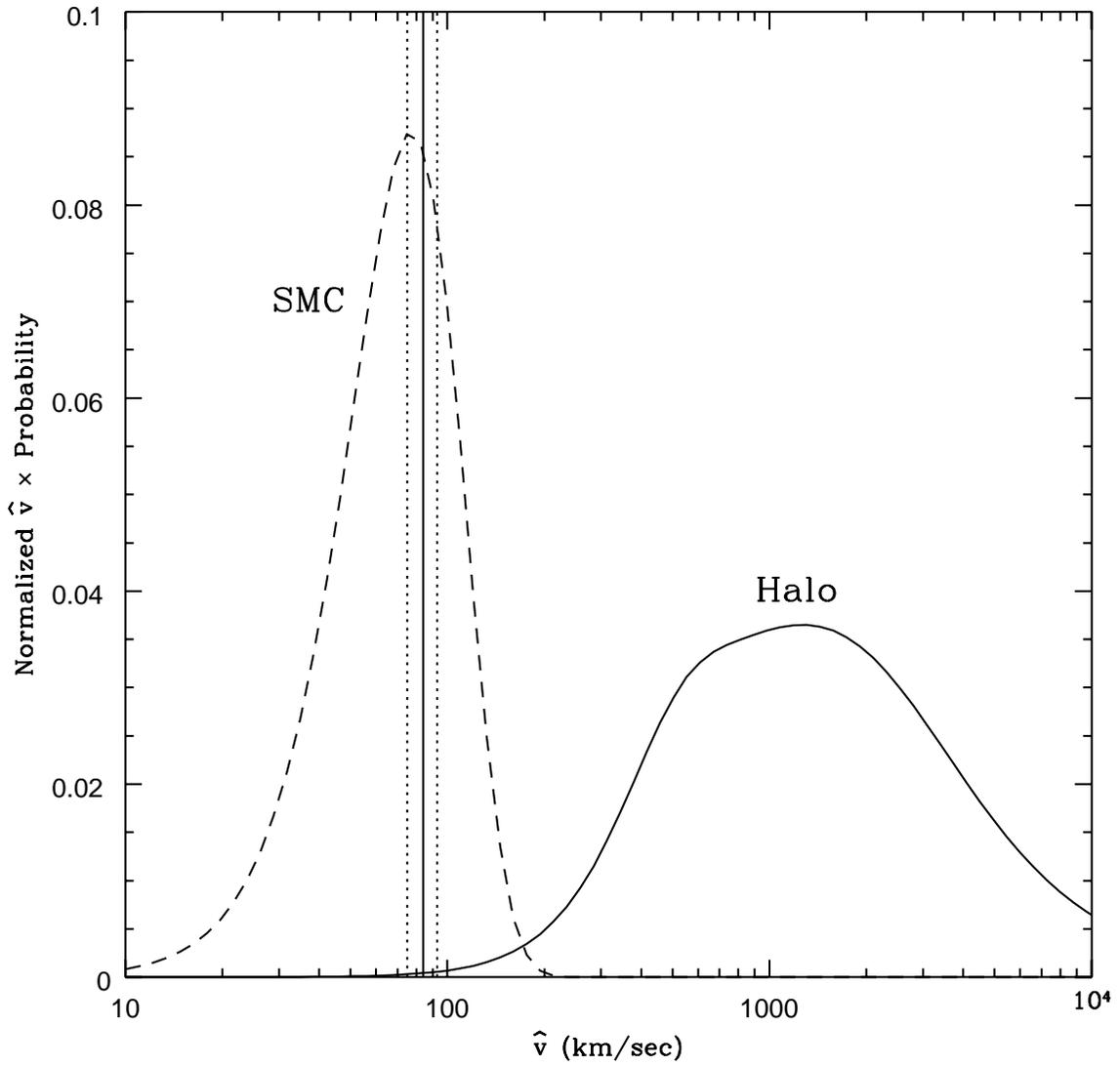}
\figcaption[f3.ps]{\label{fig-vhat}
Predicted $\vhat$ distributions for halo and SMC lenses.  The measured value
of $\vhat$ and error bars are indicated with vertical lines.}
\end{figure}

Figure~\ref{fig-vhat} shows the predicted $\vhat$ distributions as
calculated from eq.~\ref{eq-like} with the measured $\vhat$ value
indicated.  The distributions peak at $\sim 60 \kms$ for SMC lenses
and $1500 \kms$ for halo lenses.  (Note that since
Figure~\ref{fig-vhat} has a logarithmic $x-$axis, what is actually
plotted is $\vhat {\cal L}(\vhat) = d\Gamma / d\log \vhat$, so areas
under these curves represent relative probabilities.)  The peak for
SMC lenses occurs at a value larger than $30 \kms$ for several
reasons: $\vhbold$ has two components, it is the {\sl relative}
source-lens motion, and there is the factor of $\vperp$ in the event
rate per lens.  Integrating these curves we find that only $0.15\%$ of
halo lenses would have a $\vhat \le 84 \kms$, while $\sim 31\%$ of SMC
lenses would have a $\vhat$ larger than this value. {\sl Thus it is
highly probable that this binary lens is located in the SMC }.

\section{Conclusions}
\label{sec-con}

The MACHO Alert system has detected the first on--going exotic
microlensing event towards the Magellanic Clouds.  The combination of
real--time event detection and GMAN follow--up observations provided
early recognition of 98-SMC-1 as a caustic crossing binary lens, and
allowed an accurate prediction of the second caustic crossing time.
MACHO/GMAN instituted intense photometric coverage of the event
leading up to and during the second caustic crossing, allowing a
unique binary lens solution for 98-SMC-1.  Including data taken by the
EROS collaboration during the caustic crossing \cite{eros-98smc1}
allows a measurement of the lens proper motion.

Assuming a lensed source with $\rstar = 1.4\pm 0.1 \rsun$ located at
$60\kpc$, the measured crossing time of $\tstar = 0.116\pm 0.010$ days
gives an estimate of the lens velocity projected to the SMC of $\vhat
= 84 \pm 9\kms$.  This relatively slow crossing indicates the source
and lens are likely members of the same population in the SMC. The
probability for a Galactic halo lens to produce a $\vhat$ value this
small is of order $0.2\,$\%.  This conclusion is substantially
stronger than that of the EROS collaboration \cite{eros-98smc1} who
concluded that $\sim 7\,$\% of Galactic halo lenses were consistent
with their caustic crossing data: this is because they were only able
to constrain one component of $\vhat$ without a fit to the full
microlensing lightcurve. The PLANET collaboration~\footnote[4]{ PLANET
maintains information on this microlensing event at

\bf{ http://www.astro.rug.nl/$\sim$planet/MS9801.html } }
\cite{planet-98smc1} did fit the entire microlensing lightcurve, but
neither of the two fits they present appears to be consistent with the
MACHO and CTIO pre-caustic crossing data.  However, our conclusions
are qualitatively consistent with both EROS and PLANET analyses of
this event.

Since this event is likely due to a lens in the SMC, and the
previously discovered SMC event also may be due to an SMC lens
\cite{eros-smc,macho-smc}, it is interesting to discuss the measured
vs. expected optical depth for SMC self-lensing.  We have not yet
performed a careful efficiency calculation for the SMC, but with about
2.2 million stars monitored over about 5.1 years, and using SMC
sampling efficiencies of about 30\%, we estimate an observed optical
depth $\tau_{est} = 2-3 \times 10^{-7}$ for the two known SMC events.
For SMC self-lensing, \citeN{eros-smc} use a prolate ellipsoid model
aligned along the line-of-sight to predict optical depths between $1
\times 10^{-7}$ and $1.8 \times 10^{-7}$, depending upon the extent of
the SMC along the line-of-sight.  Given the small number of events,
this estimate is quite consistent with the observed optical depth.  For
completeness, we note that the predicted optical depth for halo lensing
towards the SMC is about $6 \times 10^{-7}$ for a 100\% Macho halo.

There has been one previous caustic crossing binary lensing event
detected towards the Magellanic Clouds, MACHO LMC-9, reported in
\citeN{macho-bennett96a}.  This event occurred before the MACHO Alert
system was fully functional, and was not discovered while in progress.
However, there were two observations apparently taken during the first
caustic crossing.  These yield an estimate of $\vhat \sim 20 \kms$
which is rather slow, but is still consistent with a lens residing in
the LMC if the LMC has a low velocity dispersion. If the LMC had
self-lensing optical depth large enough to explain the bulk of the LMC
microlensing events, then a high LMC velocity dispersion would also be
expected \cite{gould95}.  This would be difficult to reconcile with
the $\vhat$ estimate for LMC-9, suggesting that more complicated
caustic crossing models (such as a binary source model) should be
considered.  The real--time discovery of MACHO 98-SMC-1 has resulted
in the collection of enough data to avoid such ambiguities for this
event.

Similar follow--up observations of future Magellanic Cloud
microlensing events may also yield estimates of the lens distance
through the observation of caustic crossings or other exotic
microlensing phenomena, such as the parallax effect and binary source
effects.  A statistically significant sample of such events may help
resolve the mystery posed by the microlensing results towards the
Large Magellanic Cloud.

\acknowledgements
\section*{Acknowledgments}

We are very grateful for the skilled support of the staff and
observers at CTIO, in particular R.Schommer and M.Phillips, for
flexibility in scheduling GMAN follow--up observations.  We thank the
NOAO for making nightly use of the CTIO 0.9~m telescope possible.  We
would also like to thank the EROS collaboration for making their
caustic crossing images available, and the PLANET collaboration for
exchanges on the predicted date of the crossing while the event was in
progress.

Work performed at LLNL is supported by the DOE under contract
W7405-ENG-48.  Work performed by the Center for Particle Astrophysics
personnel is supported in part by the Office of Science and Technology
Centers of NSF under cooperative agreement AST-8809616.  Work
performed at MSSSO is supported by the Bilateral Science and
Technology Program of the Australian Department of Industry,
Technology and Regional Development.  WJS is supported by a PPARC
Advanced Fellowship.  KG acknowledges support from DOE Outstanding
Junior Investigator, Alfred P. Sloan, and Cottrell awards.  CWS thanks
the Packard Foundation for their generous support.

\clearpage


\clearpage

\begin{deluxetable}{lllllll}
\tablecaption{98-SMC-1 binary microlensing event parameters \label{tab-fpar} }
\tablewidth{0pt}
\tablehead{
        \colhead {$\t0$ \tablenotemark{a} } &
        \colhead {$\that$ (days) } &
        \colhead {$\umin$ } &
        \colhead {a } &
        \colhead {$\theta$ (rad) } &
        \colhead {$\epsilon_1$ } &
        \colhead {$\tstar$ (days) } 
}
\startdata
    14.5 
  & 149   
  & 0.043
  & 0.678
  & -0.205
  & 0.276
  & ?   \nl
    14.931   (15)
  & 147.58   (41)
  & 0.04628  (12)
  & 0.66365  (84)
  & -0.1803  (18)
  & 0.27929  (57) 
  & 0.116  (10) \nl
\enddata
\tablenotetext{a} { Date in June UT. }
\tablenotetext{} { Line 1 presents the preliminary binary lens fit
parameters announced on June 15, before the second caustic crossing.
The second line shows the current best fit binary lens parameters used
in the analysis. }
\end{deluxetable} 

\begin{deluxetable}{cccccc}
\tablecaption{98-SMC-1 binary microlensing blending parameters \label{tab-bpar} }
\tablewidth{0pt}
\tablehead{
        \colhead {${\rm CTIO~R} $ } &
        \colhead {${\rm CTIO~B} $ } &
        \colhead {${\rm EROS~R} $ } &
        \colhead {${\rm EROS~B} $ } &
        \colhead {${\rm MACHO~R} $ } &
        \colhead {${\rm MACHO~V} $ }
}
\startdata
    0.79
  & 1.00
  & 0.83
  & 0.40
  & 0.47
  & 0.56 \nl
\enddata
\tablenotetext{} { Blend fractions are in the sense of ${\rm flux_{lensed} / flux_{total}}$. }
\end{deluxetable} 

\begin{deluxetable}{ccllc}
\tablecaption{98-SMC-1 fit statistics \label{tab-fstat} }
\tablewidth{0pt}
\tablehead{
        \colhead {Passband} &
        \colhead {$\#$ Observations} &
        \colhead {$\bar{dm}$ \tablenotemark{a} } &
        \colhead {$\bar{dm}$ \tablenotemark{b} } &
        \colhead {Binary Lens Fit $\chi ^2$ \tablenotemark{c} } 
}
\startdata
CTIO R   & 84
         & ---
         & 0.041
         & 44.4     \nl
CTIO B   & 22
         & ---
         & 0.045
         & 10.4     \nl
EROS R   & 38
         & ---
         & 0.20
         & 19.5     \nl
EROS B   & 38
         & ---
         & 0.077
         & 12.2     \nl
MACHO R  & 704
         & 0.55
         & 0.10
         & 921.5    \nl
MACHO V  & 712
         & 0.49
         & 0.093
         & 763.9    \nl
\tableline
TOTAL    & 1598
         &
         &
         & 1771.9   \nl
\enddata
\tablenotetext{a} { Average photometric error, in magnitudes, before alert. }
\tablenotetext{b} { Average photometric error, in magnitudes, after alert. }
\tablenotetext{c} { The binary lens fit contains 7 global constraints,
plus 2 additional baseline constraints per passband. }
\end{deluxetable}

\clearpage

\end{document}